\documentclass[%
reprint, 
superscriptaddress,
preprintnumbers,
 amsmath,amssymb,
 aps,
prb,
citeautoscript,
floatfix, 
]{revtex4-1} 

\usepackage[caption=false]{subfig}
\usepackage{enumerate}
\usepackage{graphicx}
\usepackage{dcolumn}
\usepackage{bm}

\renewcommand{\imath}{i}

\showboxdepth=\maxdimen
\showboxbreadth=\maxdimen

\usepackage[table]{xcolor}
\usepackage[normalem]{ulem}

\newcommand{\ee}{{\rm e}}
\newcommand{\ii}{{\rm i}}

\begin{document}


\title{Synchronization of \textit{E. coli} bacteria moving in coupled wells}

\author{A. Japaridze}
\affiliation{SoundCell B.V. Delft, The Netherlands}
\affiliation{Delft University of Technology, Delft The Netherlands}
\author{V. Struijk}
\affiliation{Delft University of Technology, Delft The Netherlands}
\author{K. Swamy}
\affiliation{Delft University of Technology, Delft The Netherlands}
\author{I. E. Rosłoń}
\affiliation{SoundCell B.V. Delft, The Netherlands}
\affiliation{Delft University of Technology, Delft The Netherlands}
\author{O. Shoshani}
\affiliation{ Ben-Gurion University of the Negev, Beer-Sheva, Israel}
\author{C. Dekker}
\affiliation{Delft University of Technology, Delft The Netherlands}
\author{F. Alijani}
\affiliation{Delft University of Technology, Delft The Netherlands}
\affiliation{Corresponding author: f.alijani@tudelft.nl}

\date{\today}

\begin{abstract}
Synchronization plays a crucial role in the dynamics of living organisms, from fireflies flashing in unison to pacemaker cells that jointly generate heartbeats. 
Uncovering the mechanism behind these phenomena requires an understanding of individual biological oscillators and the coupling forces between them. Here, we develop a single-cell assay that studies rhythmic behavior in the motility of individual \textit{E.coli} cells that can be mutually synchronized. Circular microcavities are used to isolate \textit{E.coli} cells that swim along the cavity wall, resulting in self-sustained oscillations. Upon connecting these cavities by microchannels the bacterial motions can be coupled, yielding nonlinear dynamic synchronization patterns with phase slips. 
We demonstrate that the coordinated movement observed in coupled \textit{E. coli} oscillators follows mathematical rules of synchronization which we use to quantify the coupling strength. These findings advance our understanding of motility in confinement, and lay the foundation for engineering desired dynamics in microbial active matter.


\end{abstract}

\maketitle

\onecolumngrid
Life at low Reynolds numbers remains intriguing \cite{LifeAtLowRe}. Flagellum-driven motility enables bacterial cells to explore their environment, find nutrients, and avoid toxins \cite{TreeOfMotility,Chemotaxis}. Bacterial motility also provides insights into the formation of biofilms \cite{Biofilms}, bacterial swarming \cite{kearns2010field}, the propagation of infections \cite{Infection}, and has been even used as a measure to determine the efficacy of antibiotics in rapid antibiotic susceptibility testing \cite{roslon2022probing, roslon2023microwell}.

It is well-established that motile bacteria such as \textit{Escherichia coli} exhibit a random-walk behavior \cite{berg1993random,wadhwa2022bacterial}, with periods of straight locomotion (swimming phase) that alternate with moments of abrupt reorientation (tumbling phase). Near flat surfaces, however, \textit {E. coli} cells  suppress their tumbling frequency \cite{surface_swim} and follow a circular trajectory\cite{Surface_circles, lauga2016bacterial}. This sporadic periodic movement is rooted in the spatial distribution of the flagella bundle and hydrodynamic interaction with the nearby surface.  
Interestingly, it has also been demonstrated that these rotations can get weakly phase-locked to one another in dense populations and lead to large scale collective dynamics, for which the origins have have remained incompletely understood \cite{collective_oscillation_ecoli}. 
It is of interest to understand the microscopic processes behind this coordinated movement and devise effective strategies to control it.

Here, we engineer single-cell \textit {E. coli} oscillators and develop a new assay to systematically study the role of hydrodynamic interactions in their rhythmic motion. Inspired by the advancements in manipulating bacterial motility via physical boundaries \cite{surface_swim,MagneticVortex,Racetrack}, we build circular microcavities that trap single \textit{E.coli} cells from a bulk population. We show that these bacteria perform a continuous circular motion, yielding periodic oscillations over minute time scales that can be adjusted by engineering the cavity dimensions. When mutually coupling the microcavities with an interconnecting microchannel, we observe that the \textit{E.coli} bacteria can couple their swimming patterns and exhibit long periods of in-phase oscillations. From a stochastic nonlinear dynamics model, we extract the coupling strength and optimize the channels to mediate synchronized oscillations between these bacterial oscillators. The findings pave the way to engineering micro-tools for inducing controlled oscillations and synchronization in bacterial active matter \cite{aranson2022bacterial}, and provide an understanding of the microscopic origins of spontaneous order emerging among the smallest living organisms. 


We studied \textit{E.coli} (deltaCheA strain) bacteria swimming over an array of circular PDMS microcavities that had a diameter of $d$=8$\mu m$ and a depth of 2.5$\mu m$ (see Methods and Supplementary Information I). The tendency of \textit{E.coli} cells to move towards solid boundaries \cite{frymier1995three} led to cells that were continuously swimming inside the microcavities where they performed circular paths that tracked the side walls of the cavity (See Supplementary Video 1). This assay provided the possibility to sieve single cells of \textit{E.coli} from the population and study their motion in confinement.
To record the motion of single cells, we used widefield phase contrast microscopy (see Figure \ref{fig:fig1}a,b) and tracked the position of the cells relative to the cavity (see Supplementary Information II for the details). We found that \textit{E.coli} cells that settled inside the cavities swam continuously in the clockwise direction along the cavity wall. This swimming pattern with a set chirality direction can be attributed to the right-handedness of \textit{E.coli} flagella which makes these bacteria 'swim on the right-hand side' \cite{Right_handed}. 

Figure \ref{fig:fig1}c-d shows a typical trace, where a single \textit{E.coli} cell was swimming inside an 8$\mu m$ microcavity with a linear speed of $v \approx 3.5$ $\mu m/s$. It can be observed that the cell performs a harmonic motion $x+\ii y=(d/2)\ee^{\ii\varphi}$ (Figure \ref{fig:fig1}d) with a phase angle $\varphi$ that changes uniformly in time $\varphi=\omega t$, where the angular frequency is $\omega=-2v/d$ $\approx -0.9/s$ (see also Supplementary Video 2). This persistent periodic behavior was observed even up to 13 minutes in some cases (See Extended Fig. 1). Cell trapping was found to be possible for various geometries of the cavity, as cells showed clock-wise rotations also in rectangular traps and square labyrinths (See Supplementary Videos 3 and 4). The swimming pattern was also apparent in measurements conducted on cells that settled inside inverted microcavities, where \textit{E. coli} overcame gravity and exhibited rotary motion along the ceiling (See Supplementary Video 5). 
Beyond the regular circular motion, we observed significant fluctuations in the rotary motion, see 
Figure \ref{fig:fig1}d. 
To characterize the noise, we decomposed the velocity $v$ into two parts, viz., $v=\langle v\rangle+\delta v$, where the mean velocity $\langle v\rangle$ generates the periodic motion.
The velocity fluctuations $\delta v$ from the mean velocity $\langle v\rangle$ 
were characterized as zero-mean delta-correlated Gaussian noise, i.e., $\langle \delta v(t) \delta v(t+\tau)\rangle=2\sigma^2 \delta(\tau)$, where the numerical value of $\sigma$ is the standard deviation.
To quantify $\langle v\rangle$ and $\sigma$, we performed a large number of measurements ($N=291$) on single bacteria trapped in $d$=8$\mu m$ cavities. This yielded 
$\langle v\rangle$ = 6.5 $\mu m/s$ and $\sigma$ = 2.6 $\mu m/s^{1/2}$ (see Table II in Supplementary Information III for details). We note that the frequency of  the rhythmic motion we probe in different microcavities range between $0.1- 0.3$ Hz, which is smaller than the $~1$ Hz tumbling frequency of \textit{E.coli}\cite{berg1972chemotaxis}. On the other hand,  $\sigma^2$ in our measurements are much larger than the  Brownian diffusion coefficient of $~0.1 $ $\mu m^2 /s$\cite{tavaddod2011diffusion}, suggesting that the noise in our \textit{E. coli} oscillators does not only stem from Brownian motion.  

We further found that the rotational speed of the bacteria depends on the cavity size. We performed single-cell measurements on PDMS microcavities of diameters ranging from $d$=5 to 30$\mu m$. Figure \ref{fig:fig2}a shows examples of measured cell trajectories for a 7$\mu m$ and a 25$\mu m$ microcavity. Whereas cells trapped in the 7$\mu m$ microcavities were found to continuously follow the cavity wall similar to Figure \ref{fig:fig1}c, cells in the large 25$\mu m$ cavities exhibited two distinct types of dynamics (see Supplementary Video 6 and Extended Fig. 2): trapped cells were observed to have periods of clockwise spiraling within the cavity interior, i.e. without running along the cavity walls, which alternated with periods of swimming clockwise along the cavity wall.
We observed (see Extended Fig. 3) that the bacterial activity in 7$\mu m$ cavities was almost entirely concentrated at the cavity edge (99\%), while this was reduced to 77\% in the 25$\mu m$ microcavities. 
For the 7$\mu m$ microcavities, we found that $v=5.6 \pm 2.2$ $\mu m/s$ (mean $\pm$ sd), while we observed an almost twice higher speed of $v=10 \pm 3.3$ $\mu m/s$ (mean $\pm$ sd) for the $25$ $\mu m$ microcavities.
Given this sizeable difference, we quantified the speed of single-cell rotary motion for a range of confinement sizes (Figure \ref{fig:fig2}c). We found that cells slowed down significantly when driven in more strongly confined cavities, reducing their average speed from $~$10$\mu m/s$ (which is 75\% of the measured speed of a cell on a free surface) in cavities with diameter $d\ge$14$\mu, m$ to 5$\mu m/s$ for $d\le$6$\mu m$. 
We speculate that this drop-off in speed occurs as the cavities become smaller than the size of an \textit{E.coli} cell with its flagella \cite{depamphilis1971purification}. Interestingly, the cell residence time, i.e., the time cells remain trapped inside the circular well, was about 1 $min$ on average, independent of diameter (See Extended Fig. 4).

Next, we investigated the motion of a \textit{pair} of \textit{E. coli} cells in two neighboring microcavities. We noticed that generally two cells in adjacent wells of the same size did not show signs of coupled dynamics (see Supplementary Video 7). The phase difference between them ran freely as a function of time - despite a mutual distance that was only equal to the cavity diameter (~8$\mu m$). This indicates that  
hydrodynamic couplings through the bulk fluid beyond the cavities were insufficient to induce synchronisation. 
However, when we connected the cavities with a microchannel (see Figure \ref{fig:fig3}a), we strikingly observed that two bacteria move in unison (see Fig. \ref{fig:fig3}b and Supplementary Video 8). The channel allowed the fluid exchange between the microcavities but was too narrow to allow bacteria swim through \cite{mannik2009bacterial}. 
To experimentally detect the correlated motion of the bacteria, we defined a time-dependent correlation function $\rho\equiv\cos\varphi $ \cite{arenas2006synchronization} in which $\varphi=\varphi_2-\varphi_1$ is the measured phase difference between the two neighboring bacteria as a function of time. 

Figure \ref{fig:fig3}b shows the result of one such experiment where two bacterial cells moved synchronously in connected microcavities ($d$= 8 $\mu m$, $c_w=0.5\mu$m, $c_\ell=0.5\mu$m). From Figure \ref{fig:fig3}b, it can be observed that the motion of cells is partially phase-locked, and the coupled dynamics involves distinct transitions from calmer epochs where $\rho\approx1$, signalling perfect synchronization, to periods of large modulations where $\rho\neq1$. This intriguing observation is reminiscent of the slow-fast dynamics observed theoretically at the onset of synchronization \cite{strogatz2004sync}. 
To explore the possibility of modulating the coupling strength, we designed channels of different length ($c_\ell$) and width ($c_w$) and generated histograms of the phase difference (modulo $2\pi$) between neighboring cells for a large number of connected microcavities with $d$= 7 $\mu m$ (see Figure \ref{fig:fig3}c). We observed that the histograms became sharper for channels that were shorter, indicating a greater likelihood of achieving synchronization. 

To quantify the strength of the coupled motion, we fitted our data with the Adler equation
$\dot\varphi = \Delta \omega - k \sin\varphi+\xi(t)$ \cite{strogatz2018nonlinear}.
Here, $\Delta \omega$ represents the frequency mismatch $\Delta \omega=\omega_2 - \omega_1$ between two adjacent cells, $k$ is the coupling parameter, and $\xi(t)$ denotes zero-mean delta-correlated Gaussian noise, i.e., $\left< \xi(t) \right>=0$ and $\left< \xi(t)\xi(t+\tau) \right> =4 (\sigma/d)^2 \delta(\tau)$, where $\sigma$ is measured from single-cell velocity fluctuations $\delta v$ 
and the factor 4 comes from the assumption that noise in the system is uncorrelated (see Supplementary Information III). 
The Adler equation, which describes the motion of an overdamped particle sliding on a washboard potential\cite{gitterman2008noisy}, is one of the simplest models for studying synchronization that can be derived from heuristic arguments  (see Supplementary Information III) \cite{pikovsky2001universal}. For the coupled motion of bacteria this equation can be also derived from analysis of fluid-rotor interaction in Stokes flow, where the hydrodynamic coupling is mediated by the normal viscous force transmitted through the channel (See Supplementary Information IV).

To obtain the coupling parameter $k$ from our data, we linearized the Adler equation around the stable phase difference, which is approximately $\varphi\approx0$ (Figure \ref{fig:fig4}c), and calculated the variance $\langle \varphi^2\rangle=(2/k) (\sigma/d)^2$, see Methods for details. Since we know $\sigma$ for the microcavities, and experimentally measured $\langle \varphi^2\rangle$, we could obtain the coupling parameter $k$ for different channel dimensions (see Methods for details). As expected, we found that shorter channels yielded an increased $k$ while the coupling got lost with increasing $c_\ell$ such that for $c_\ell= 2 \mu m $, only mild coupling could be observed. We also noticed that the dependence of $k$ on $c_w$ was minor since changing it from $0.5\mu$m to $0.9\mu$m did not yield a noticeable increase in the coupling strength (see Table I). These findings are consistent with a minimalistic model based on Stokes flow which highlights  the presence of channel for the bacteria to sync and suggests that the hydrodynamic coupling $k\propto c_w/ c_\ell^2$ (See Supplementary Information IV). 
Experimental verification of the dependence of coupling  on $c_w$ was difficult as increasing the channel width beyond $0.9\mu$m caused the bacteria to swim through.

Using the Adler equation and the experimentally measured average rate of the phase difference $\langle\dot\varphi\rangle$, we also estimated the frequency mismatch $\Delta\omega$ of the pair of \textit{E. coli} cells moving in connected cavities (see Methods). Figures \ref{fig:fig4}a and \ref{fig:fig4}b show Shapiro-like plateaus obtained by fitting the Adler equation \cite{pikovsky2001universal} to our data. As expected, we observed wider plateaus for shorter channels. In Figure \ref{fig:fig4}c,d we report a typical correlation function $\rho$ and phase difference $\varphi$ when $k>\Delta\omega $ and thus within the synchronization plateau. Supplementary Video 9 shows the same synchronized \textit{E. coli} oscillators. It can be observed that the phase difference remains in the vicinity of zero with small-amplitude fluctuations that are induced by noise. In Figure \ref{fig:fig4}e,f we also show results from the numerical integration of the noisy Adler for the same experimental configuration in which we used the estimated parameter of the experiments ($\sigma$, $k$, and $\Delta\omega$). These results demonstrate that single \textit{ E. coli} oscillators display nonlinear dynamics behaviour in confinement through hydrodynamic forces. Interestingly, the synchronized motion is well reproduced by the simplest mathematical model of synchronization i.e, the Adler equation.

Summing up, we presented a platform comprising arrays of circular microcavities to study the motion of coupled \textit{E. coli} cells in confinement. The bacteria were found to exhibit clock-wise rotations along the cavity walls that could be sustained for over hundred cycles. 
By devising sets of microcavities that were pairwise connected by channels, we showed that \textit{E.coli} cells can coordinate their motion to  neighboring cells, thus exhibiting coupled oscillatory behavior. 
The channels could be engineered to induce strong enough hydrodynamic coupling that led to rich nonlinear dynamic phenomena, including slow-fast dynamics and synchronous oscillations in the presence of noise. The microcavities could be even designed 
to concentrate and study the synchronous motion of multiple cells in a circular cavity, analogous to runners on a race track (see Supplementary Video 10).
Previous experimental studies already hinted at weak synchronization of \textit {E. coli} cells from unknown origins in dense populations \cite{collective_oscillation_ecoli}. Highly concentrated suspensions of \textit {Bacillus subtilis} and \textit {Pseudomonas aeruginosa} also showed emergent nonlinear dynamic behaviors including propagating spiral waves \cite{liu2024emergence}, active turbulence \cite{dunkel2013fluid} in bulk, and self-organization in confinement \cite{MagneticVortex}. Our single-cell data expand on these intriguing observations, elucidating the role of hydrodynamic forces in the generation of coupling between adjacent micro-swimmers. Furthermore, it shows that spontaneous order can be engineered and controlled at the level of single cells. 

Synchronization is important in biology, physics, and engineering across different time and length scales, from planetary resonances in the solar system \cite{strogatz2004sync}, to the synchronous flashing of fireflies \cite{buck1968mechanism}, or even the spontaneous clapping of the audience in a theater \cite{Applause}. Here, we report evidence of synchronization between single bacteria, taming their random motion and engineering order out of their chaotic dynamics.
Our results support theoretical works on hydrodynamically coupled active components \cite{Carpet_of_Rotors}, and call for further studies to understand, control, and unravel the stochastic nonlinear phenomena generated by micro-swimmers in confinements. 
By shape optimization of the microcavities and channels, to enhance the coupling strength and suppress noise, we envision that bacterial oscillators can evolve into large arrays of synchronous microbial matter with adjustable couplings, paving the way to controllable micro-organism-based oscillator networks and confined active matter \cite{gompper20202020}.

\section{Methods}\label{sec:methods}
\subsection{Sample preparation}
\paragraph{Cell culture}
Smooth sailing \textit{Escherichia coli} (cheA strain) were grown in Lysogeny broth (LB). A monoclonal colony was diluted in 5ml LB and left to grow overnight at 30$^{o}C$ while gently stirred in an incubator. On the day of the experiments, an 50$\mu l$ aliquot was diluted 1:100 in volume into 5ml LB and grown for 2.5 h at 30$^{o}C$ to mid-exponential phase. A subsample of this culture was again diluted typically ~1:10 in fresh LB to finally reach an optical density (OD600) of ~0.05. 

\paragraph{Mold preparation}
Molds were prepared from 500 micron thick silicon wafers with a 285nm thick layer of silicon dioxide. The wafers were patterned by electron beam lithography followed by reactive ion etching, resulting in a feature height of 2.5 micron for PDMS casting. 

\paragraph{PDMS preparation}
Pre-polymer PDMS was mixed with curing agent in 4:1 ratio (2g PDMS to 0.5g curing agent), and stirred manually with the tip of a sterile pipette. The mixture was then degassed in a vacuum chamber for 10-15 min, until all air bubbles had disappeared. Using the tip of a pipette, a small amount of PDMS was carefully dipped in the center of one of the structures of the patterned wafer (see Supplementary Information I). Then, a 1mm thick, 22x22mm glass slide was placed on the droplet, spreading it over the patterned wafer structure. The assembly was subsequently baked at 90$^{o}C$ for 2.5h in an oven. Prior to baking, the silicon wafer was pre-treated with Tridecafluoro-l,l,2,2-tetrahydrooctyl)trichlorosilane (FTS) to ensure that the PDMS mould could be easily released after it had been cured \cite{depalma1989friction}. 
With a sharp sterile razor blade, the glass slide was carefully lifted from the wafer and stored till the day of the experiment. The PDMS stuctures were eventually plasma treated at 20W power and 60mTorr oxygen chamber pressure for 30s in order to make them hydrophyllic. 

\paragraph{Imaging}
Measurements were carried out using an inverted microscope (Nikon Ti) under bright-field illumination, through a 100× oil-immersion objective. Recordings typically had 2min duration at frame acquisition interval of 0.23s. For imaging, about 35 $\mu l$ cell culture was pipetted onto a prepared PDMS sample and placed on a coverslip holder. Then, with parafilm we placed a sample cover over the PDMS sample, creating a sealed chamber (see fig \ref{fig:fig1}a). The assembly was placed inside a closed microscopic chamber maintained at 30$^{o}C$. 

\subsection{Statistics of the noisy Adler equation}
\label{subsec:stat_adler}
The probability density of the phase difference $w(\varphi,t)$ of the Adler equation obeys the following Fokker–Planck equation\cite{stratonovich1967topics}
\begin{equation}
\frac{\partial w}{\partial t}= -\frac{\partial}{\partial\varphi} \left[(\Delta\omega-k\sin\varphi)w\right]+4\left(\frac{\sigma}{d}\right)^2\frac{\partial^2w}{\partial\varphi^2},
\label{eq:FPE}
\end{equation}
and expresses the conservation of probability $\partial w/\partial t+\partial J/\partial\varphi=0$, where 
\begin{equation}
J=(\Delta\omega-k\sin\varphi)w-4\left(\frac{\sigma}{d}\right)^2\frac{\partial w}{\partial\varphi}
\label{eq:FPE_ss}
\end{equation}
is the probability current. We look for a stationary (time-independent) probability density $w_{\rm ss}$, where $\partial w_{\rm ss}/\partial t=0$. The stationary solution of Eq. \eqref{eq:FPE} is $2\pi$-periodic in $\varphi$, and therefore, we can write it in terms of its Fourier expansion\cite{pikovsky2001universal} 
$w_{\rm ss}=\sum_{-\infty}^{\infty}W_n\ee^{\ii n\varphi}$. From the normalization condition $\int_0^{2\pi}w_{\rm ss}(\varphi)d\varphi=1$, we find that $W_0 =(2\pi)^{-1}$, and from the reality of $w_{\rm ss}$, we deduce that $W_{-n} =W^*_{n}$, where $W^*_{n}$ is the complex-conjugate of $W_{n}$. Furthermore, by integrating Eq. \eqref{eq:FPE_ss} with respect to $\varphi$ from 0 to $2\pi$, we find that
\begin{align}
2\pi J=\int_0^{2\pi}(\Delta\omega-k\sin\varphi)w_{\rm ss}(\varphi)d\varphi-4\left(\frac{\sigma}{d}\right)^2[w_{\rm ss}(2\pi)-w_{\rm ss}(0)]\nonumber\\
=\int_0^{2\pi}(\Delta\omega-k\sin\varphi)w_{\rm ss}(\varphi)d\varphi\equiv\langle\Delta\omega-k\sin\varphi\rangle=\langle\dot\varphi\rangle.
\label{eq:av_phi_dot}
\end{align}
Substitution of the Fourier expansion into Eq. \eqref{eq:FPE_ss}, yields the following tridiagonal system of equations 
\begin{equation}
-\left( \frac{4\ii n\sigma^2}{d^2}-\Delta\omega\right)W_n-\frac{\ii k}{2}(W_{n-1}-W_{n+1})=J\delta_{n,0}.
\label{eq:tridaig}
\end{equation}
From Eq. \eqref{eq:tridaig}, we find the following relation for $n>0$, $$\frac{W_n}{W_{n-1}}=\left[-\left(\frac{8n\sigma}{kd}+\frac{2\ii\Delta\omega}{k}\right)+\frac{W_{n+1}}{W_n}\right]^{-1}.$$
Setting $n=1$, and using continuous fractions, we obtain the following rapidly converging solution\cite{risken1996fokker}
$$W_1=\frac{(2\pi)^{-1}}{-\left(\frac{8\sigma^2}{kd^2}+\frac{2\ii\Delta\omega}{k}\right)+\frac{1}{-\left(\frac{16\sigma^2}{kd^2}+\frac{2\ii\Delta\omega}{k}\right)+\frac{1}{-\left(\frac{24\sigma^2}{kd^2}+\frac{2\ii\Delta\omega}{k}\right)+...}}}.$$
By setting $n=0$ in Eq. \eqref{eq:tridaig}, we find that $J=\Delta\omega/(2\pi)-k\Im\{W_1\}$, and therefore, using Eq. \eqref{eq:av_phi_dot}, we get $\langle\dot\varphi\rangle=\Delta\omega-2\pi k\Im\{W_1\}.$
\\\textbf{\textit{Estimation of the coupling parameter:}}
\\To estimate the coupling parameter, we linearize Adler equation around $\varphi=0$, and obtain
\begin{align}
  \dot\varphi=-k\varphi+\xi(t).
  \label{eq:lin_adler}
\end{align}
Eq. \eqref{eq:lin_adler} can be readily integrated to yield
\begin{align}
 \varphi(t)&= \varphi(0)e^{-k t}+\int_0^t\xi(\tau)e^{k(\tau-t)}d\tau,
\label{eq:phi(t)}
\end{align}
with a rapidly vanishing mean value $\langle\varphi\rangle=\varphi(0)e^{-k t}$. Thus, at steady-state, where $\langle\varphi\rangle=0$, the variance is given by
\begin{align}
 \langle\varphi^2\rangle&= \int_0^t\int_0^te^{k(\tau-t)}e^{k(s-t)}\langle\xi(\tau)\xi(s)\rangle d\tau ds=4\left(\frac{\sigma}{d}\right)^2\int_0^te^{2k(\tau-t)}d\tau,
\label{eq:phi(t)2}
\end{align}
which yields the steady-state variance $\langle\varphi^2\rangle=2\sigma^2/(kd^2)$, and can be used to estimate the coupling parameter $k=2\sigma^2/(\langle\varphi^2\rangle d^2)$. Using the measured noise intensity $2(\sigma/d)^2$ and the variance of the locked phase difference, we obtain the following estimations for the coupling parameter
\begin{table}[h]
  \centering
  \begin{tabular}{c|c|c|c} 
  \hline
     $k~\rm (rad/s)$ & $c_\ell=0.5~\rm \mu m$ & $c_\ell=1~\rm \mu m$ & $c_\ell=2~\rm \mu m$\\ \hline
     
   $c_w=0.5~\rm \mu m$ & 0.26 & 0.17 & 0.13\\ \hline
    $c_w=0.7~\rm \mu m$ & 0.38 & 0.18 & 0.13 \\ \hline
    $c_w=0.9~\rm \mu m$ & 0.31 & 0.19 & 0.14\\ \hline
  \end{tabular}
  \caption{Model-based estimation for the coupling parameter as a function of the channel width and length.}
  \label{table:k}
\end{table}

\section*{Data Availability}
The datasets generated during and/or analysed during the current study are available from the corresponding author. The data is fully represented in the figures as plotted lines. 

\section*{Bibliography}
 \bibliography{bibliography}
 \bibliographystyle{naturemag}
 
 \section*{Acknowledgments}
Financial support was provided from the European Union’s Horizon 2020 research and innovation programme under ERC starting grant ENIGMA (802093), ERC Consolidator grant NCANTO (101125458), ERC PoC GRAPHFITI (966720), and the ERC Advanced Grant LoopingDNA (883684). OS is supported by BSF under Grant No. 2018041, by ISF under Grant No. 344/22, and by the Pearlstone Center of Aeronautical Engineering Studies at Ben-Gurion University of the Negev. The authors thank Bertus Beaumont for the kind gift of the \textit{Escherichia coli} strain. F.A. and K.S. acknowledge fruitful discussions with Dr. Daniel Tam from Delft University of Technology. Schematics in Figure 1a were created with Biorender.com
 
 \section*{Author Contributions Statement}
A.J., and F.A. conceived the idea. A.J. and V.S. collected the data and performed the experiments. I.E.R fabricated the silicon traps. A.J, and V.S. performed the bacterial manipulation. O.S. formulated the theoretical modeling, conducted analytical and numerical analysis, and performed the fitting with the experimental data. K.S. developed the hydrodynamic model. A.J. F.A. and V.S. designed the experiments. The project was supervised by F.A, C.D. All authors contributed to the data analysis, interpretation of the results, writing of the manuscript, with the main contribution from FA.

 \section*{Competing Interests Statement}
Employment or leadership: A.J., I.E.R; SoundCell B.V. Consultant or advisory role: F.A.; SoundCell B.V. The authors declare no further competing interests.

\newpage
\begin{figure}[h]
\centering
\includegraphics[width=0.88\textwidth]{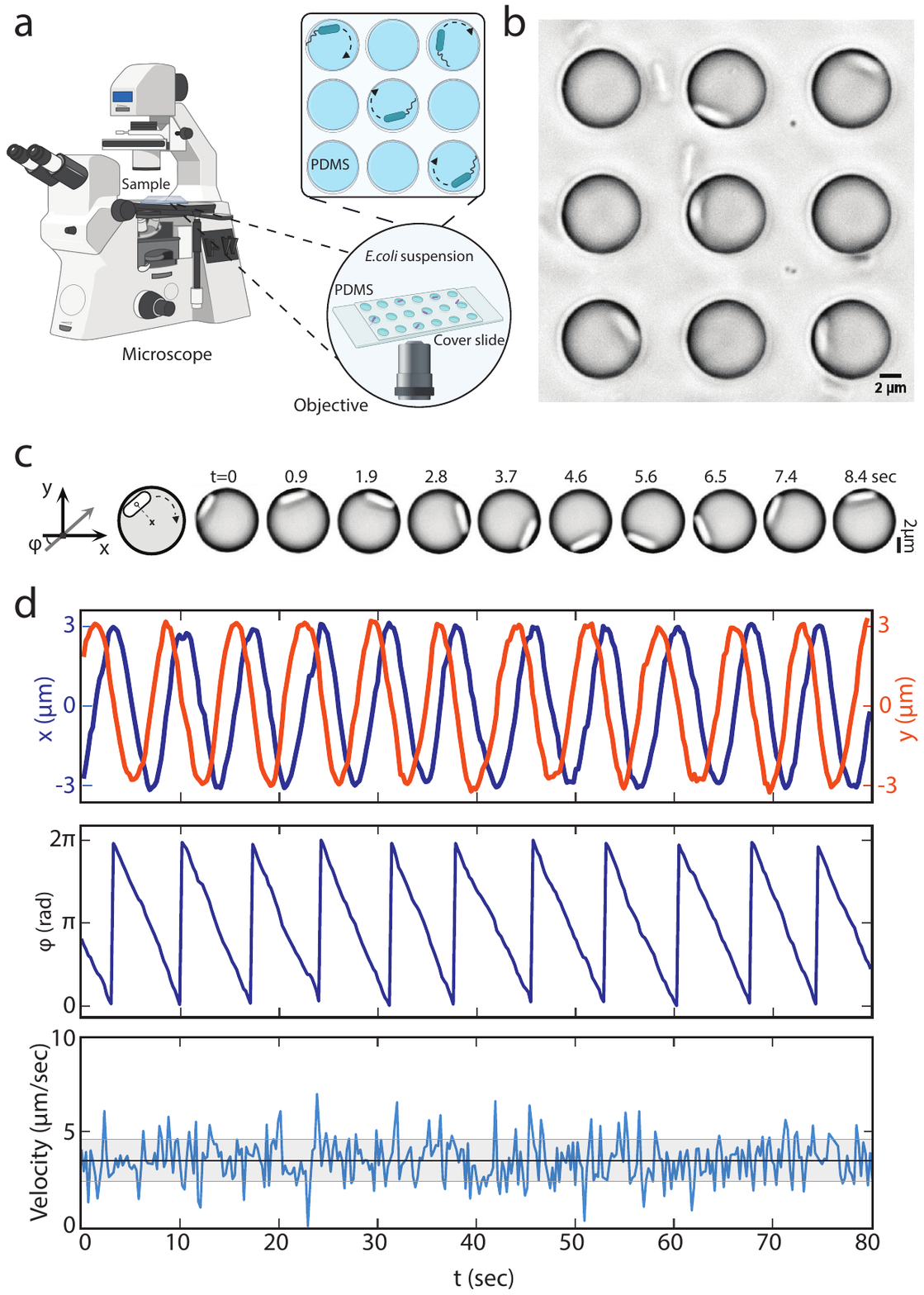}
 \caption{\textbf{Measuring cell motility in confinement.} \textbf{a} Schematic of microscope setup for phase contrast imaging. \textbf{b} Image of \textit{E.coli} cells trapped in microcavities. \textbf{c} Motion of a cell inside a microcavity (8$\mu m$ diameter). Left schematic depicts the coordinate system. Right images show a time sequence of cell positions. \textbf{d} Cell coordinates and velocity. Top traces indicate x (in blue) and y (in orange) coordinates versus time; middle shows the cell phase angle; bottom show the tangential velocity, with the black line indicating the mean speed and the grey area the standard deviation.}
  \label{fig:fig1}
\end{figure} 
\newpage

\begin{figure}[h]
\centering
\includegraphics[width=0.95\textwidth]{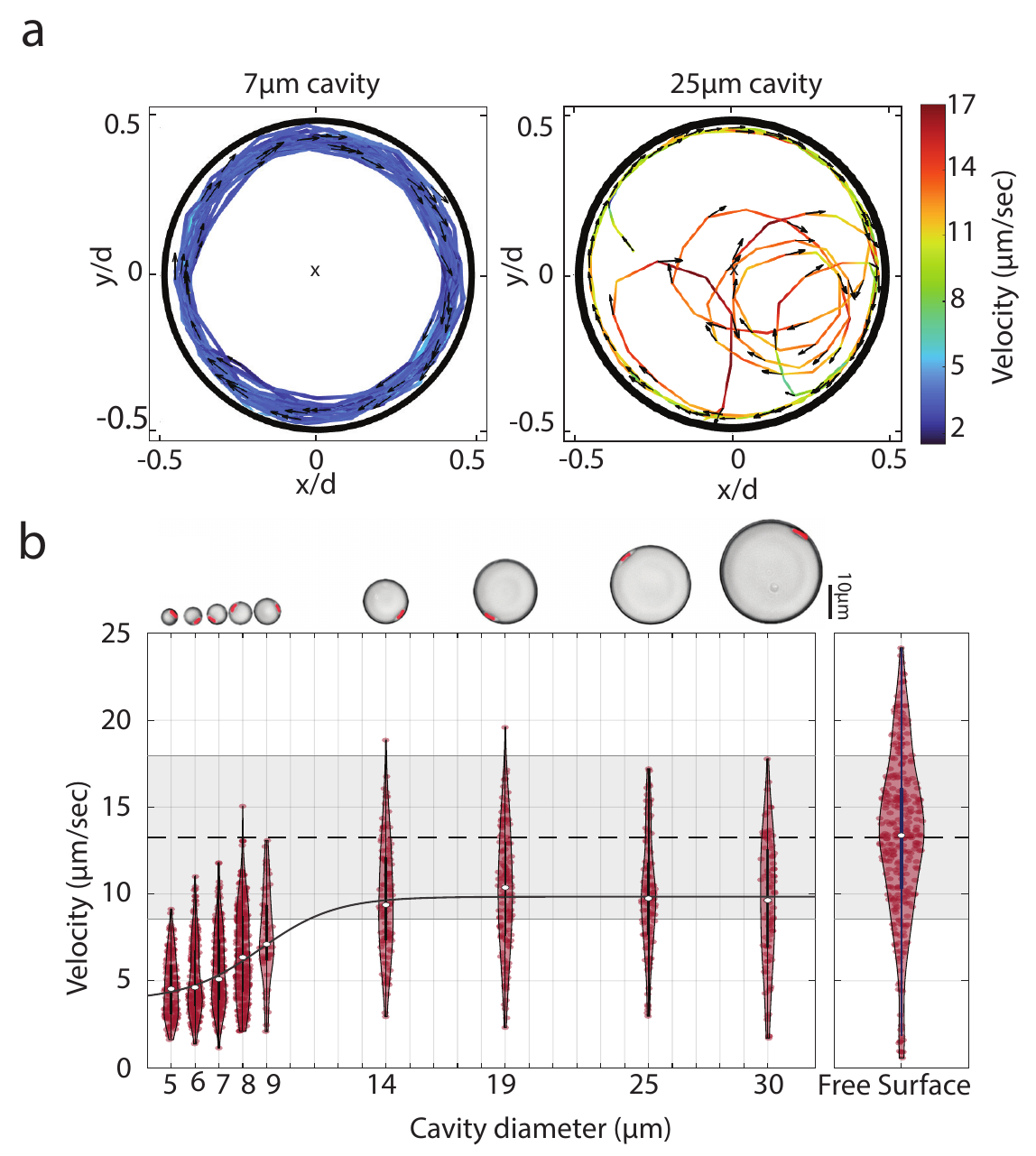}
\caption{\textbf{Influence of confinement size on \textit{E.coli} motion. }\textbf{a} Two measured cell trajectories in a 7$\mu m$ (left panel) and 25$\mu m$ (right panel) cavity. Arrows indicate swimming direction; trajectory color indicates velocity (see scale on the right). 
\textbf{b} Velocity distributions of \textit{E.coli} as a function of cavity diameter. Median values are fit with a smooth curve shown with the black line. Right panel is the distribution of speed of cells swimming over a free surface. Its the mean value is indicated by the dotted line with the grey shade indicating the standard deviation.The number of measurements are $N=154$ ($d$=5 $\mu m$), $N=217$ ($d$=6 $\mu m$), $N=226$ ($d$=7 $\mu m$), $N=291$ ($d$=8 $\mu m$), $N=46$ ($d$=9 $\mu m$), $N=104$ ($d$=14 $\mu m$), $N=125$ ($d$=19 $\mu m$), $N=100$ ($d$=25 $\mu m$), $N=83$ ($d$=30 $\mu m$), $N=304$ (free surface). }
  \label{fig:fig2}
\end{figure}

\newpage

\begin{figure}[h]
\centering
\includegraphics[width=1\textwidth]{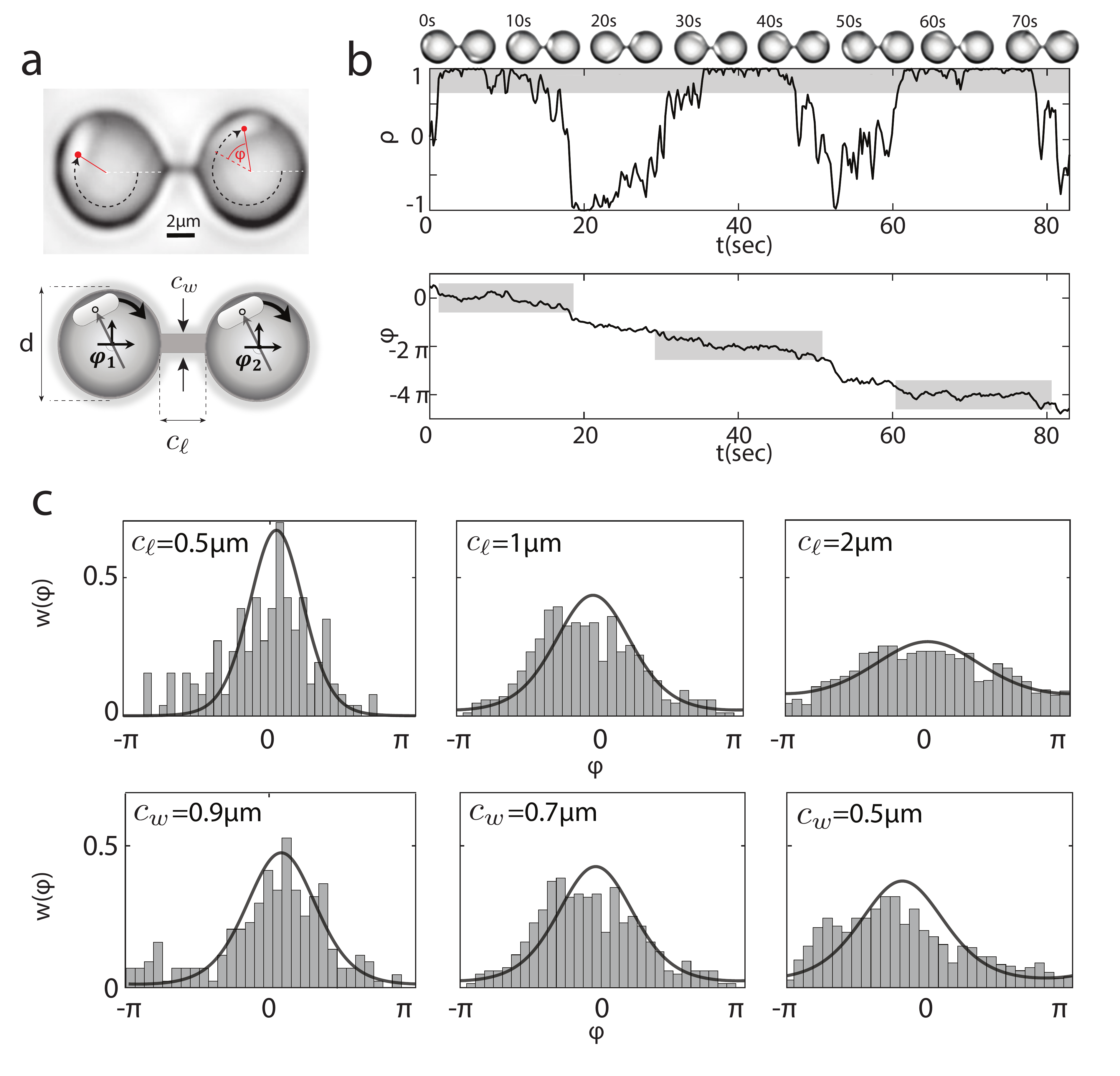}
  \caption{\textbf{Coupled \textit{E. coli} oscillators} \textbf{a} Two identical circular microcavities of diameter $d$ are connected with a channel of length $c_\ell=$ 0.5,1,2 $\mu m$ and width $c_w=$ 0.5,0.7,0.9 $\mu m$ . \textbf{b} Strong coupling is observed for the bacterial motions for $c_w=0.5\mu$m, $c_\ell=0.5\mu$m, where $\rho$ oscillates non-periodically with 
  $\langle\rho\rangle\neq0$. The phase difference $\varphi$ shows slow-fast dynamics analogous to the motion of an overdamped particle with an alternating slow-fast speed on a washboard potential. Here, grey shaded windows show the phase-locked regions. \textbf{c} Probability distribution of the phase difference $\varphi$. 
  Solid curves are fits of von Mises distribution\cite{jammalamadaka2001topics} (wrapped normal distribution). In the top panels, $c_w=0.7~\rm\mu m$ and the number of data points $(n)$ is $n$=1071 $(c_\ell=0.5~\rm\mu m)$, $n$=938 $(c_\ell=1~\rm\mu m)$, and $n$=823 $(c_\ell=2~\rm\mu m)$. In the bottom panels, $c_\ell=1~\rm\mu m$ and $n$=1224 $(c_w=0.9~\rm\mu m)$, $n$=938 $(c_w=0.7~\rm\mu m)$, and $n$=972 $(c_w=0.5~\rm\mu m)$.}
  \label{fig:fig3}
\end{figure} 

\newpage

\begin{figure}[h]
\centering
\includegraphics[width=0.9\textwidth]{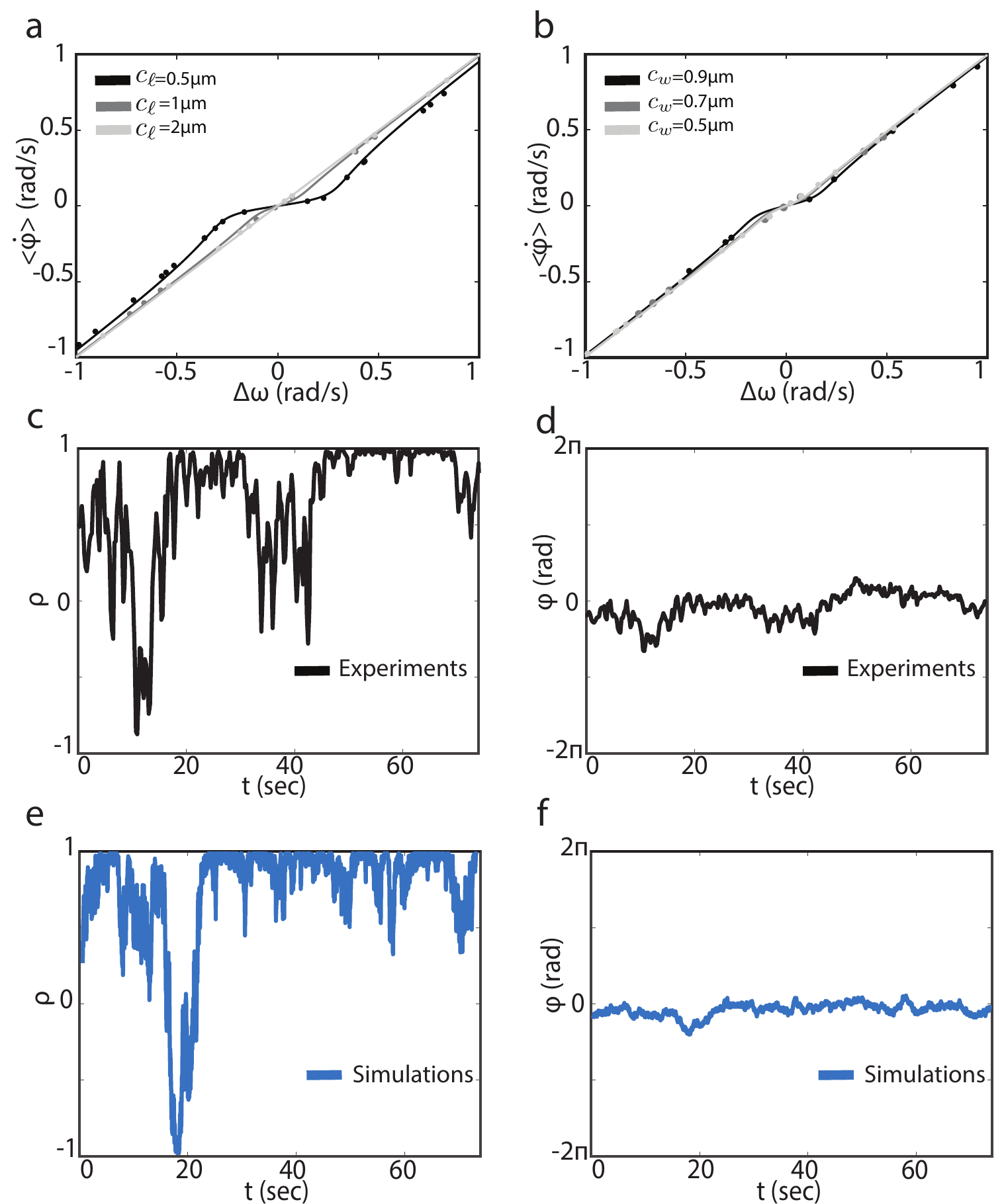}
\caption{\textbf{Synchronization region in connected microcavities.} The synchronization range for 
 fixed channel width $c_w=0.7~\rm\mu m$ but different channel lengths in \textbf{a} and for varying channel width $c_w$ but fixed channel length $c_\ell=1~\rm\mu m$ in \textbf{b}. Solid curves in (a) and (b) are the theoretical predictions, and the dots are experimental measurements with model-based estimation for the frequency mismatch $\Delta\omega$. The number of measurements $(N)$ in each curve is: $N$=26 $(c_\ell=0.5~\rm\mu m)$, $N$=22 $(c_\ell=1~\rm\mu m)$, $N$=15 $(c_\ell=2~\rm\mu)$, $N$=25 $(c_w=0.9~\rm\mu m)$, $N$=22 $(c_w=0.7~\rm\mu m)$, and $N$=8 $(c_w=0.5~\rm\mu m)$. \textbf{c, d} Synchronized motion of coupled \textit{E. coli} ($c_w=0.5\mu$m, $c_l=0.5\mu$m), where $\rho=\langle\rho\rangle\approx1$, and hence, the phase difference $\varphi$ is fully locked with the exception of noise-induced fluctuations. \textbf{e, f} numerical simulations of the noisy Adler  integrated by the Euler–Maruyama method\cite{higham2001algorithmic} for the experimental conditions of (c) and (d) 
, and extracted parameters $k=0.26$ rad/s, $\sigma=2.2$ $\mu m/s^{1/2}$, and $\Delta \omega= -0.0025$ rad/s.}
  \label{fig:fig4}
\end{figure}

\newpage
\renewcommand{\figurename}{Extended Fig.}
\setcounter{figure}{0}
\begin{figure}[h]
\centering
\makebox[\textwidth][c]{
\includegraphics[width=1\textwidth]{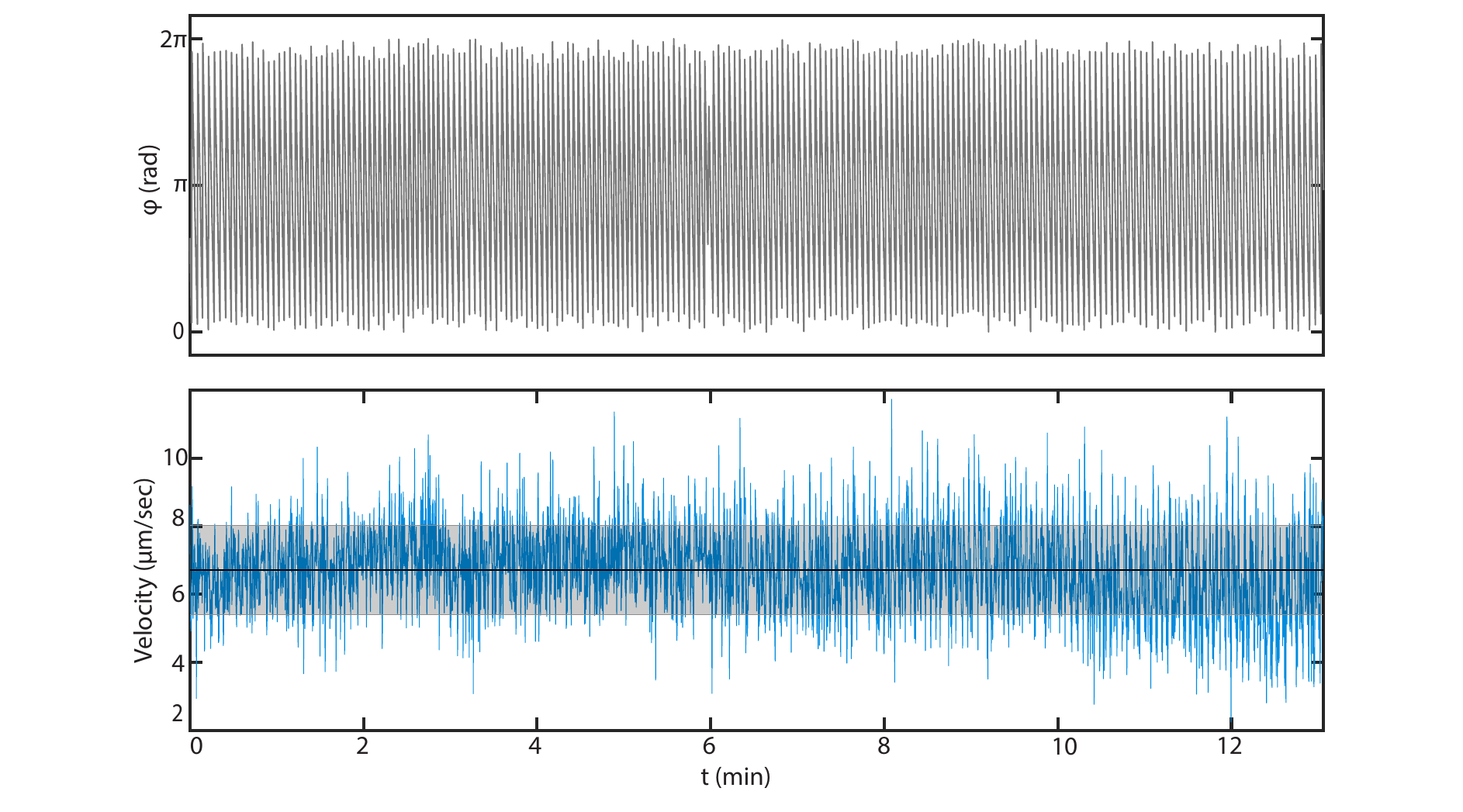}
}
\caption{\textbf{Measurement of cell motion (relative to the cavity center) inside an 8$\mu m$ microcavity}. The top panel displays cell phase angle, and the bottom panel cell velocity, with the mean speed (black line) and the standard deviation (grey shade). Trapping was sustained for over 13min, during which the bacterium was observed to traverse the cavity, performing clockwise rotations.}
\label{fig:Long trapping}
\end{figure}

\newpage

\begin{figure}[h]
\centering
\makebox[\textwidth][c]{
\includegraphics[width=1\textwidth]{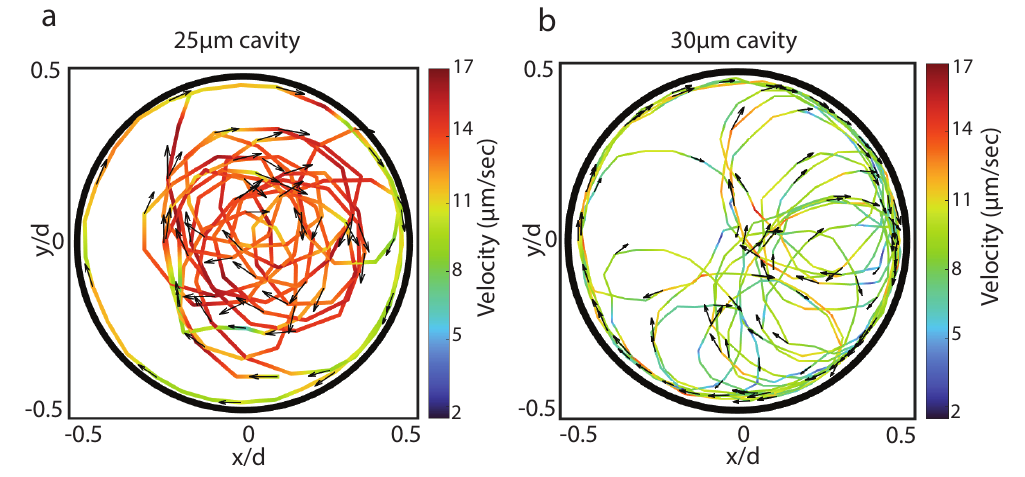}
}
\caption{ \textbf{Cell trajectories of trapped \textit{E.coli} in large cavities}. The position and velocity maps of the single-cell are displayed for (a) 25$\mu m$ cavity, and (b) 30$\mu m$ cavity. The tracking of the cell inside the cavity was conducted with ImageJ software.}
\label{fig:Extended_largewell_trajectories}
\end{figure}

\newpage

\begin{figure}[h]
\centering
\makebox[\textwidth][c]{
\includegraphics[width=1\textwidth]{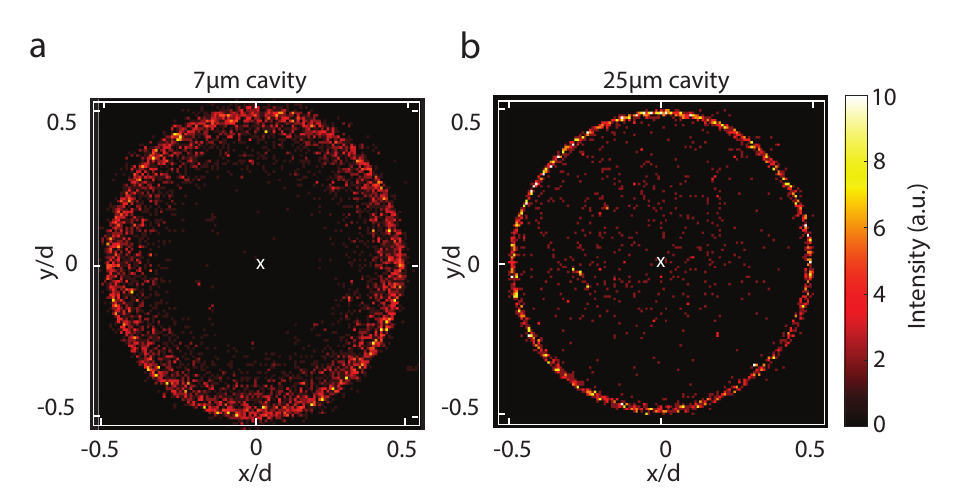}
}
\caption{ \textbf{Distribution of \textit{E.coli} motion in confinement}. \textbf{a} Averaged heat map of \textit{E.coli} motion in 7$\mu m$ microwell from $N=226$ measurements, \textbf{b} Averaged heat map of \textit{E.coli} motion in 25$\mu m$ microwell, from $N=100$ measurements.}
\end{figure}

\begin{figure}[h]
\centering
\includegraphics[width=0.9\textwidth]{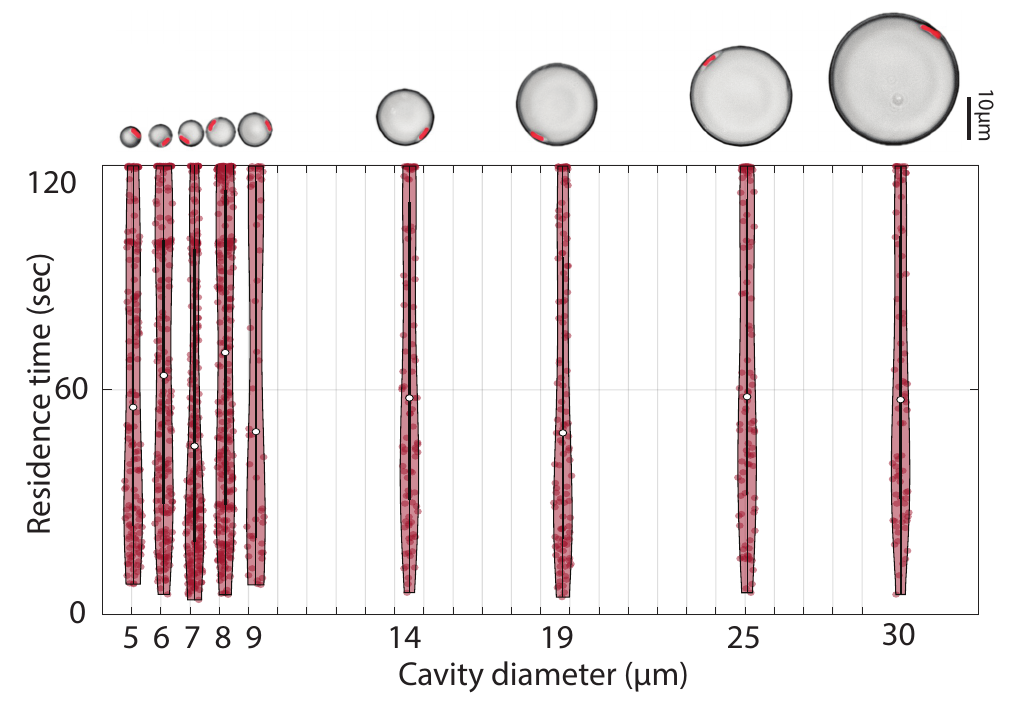}
\caption{\textbf{Residence time distribution of \textit{E.coli} in confinement.} The trapping time appears to be independent of the microcavity diameter. The recordings were cut-off at the 2$min$ mark. The number of measurements per microcavity is as in Fig. 2.}
\end{figure}

 \end{document}